# Hierarchical group dynamics in pigeon flocks

Máté Nagy[1], Zsuzsa Ákos[1], Dora Biro[2] & Tamás Vicsek[1,3]

[1]*Department of Biological Physics, Eötvös University, Pázmány Péter sétány 1A, H-1117, Budapest, Hungary.* [2]*Department of Zoology, University of Oxford, South Parks Road, Oxford OX1 3PS, UK.* [3]*Statistical and Biological Physics Research Group of HAS, Pázmány Péter sétány 1A, H-1117, Budapest, Hungary.*

**Animals that travel together in groups display a variety of fascinating motion patterns thought to be the result of delicate local interactions among group members[1-3]. Although the most informative way of investigating and interpreting collective movement phenomena would be afforded by the collection of high-resolution spatiotemporal data from moving individuals, such data are scarce[4-7] and are virtually non-existent for long-distance group motion within a natural setting because of the associated technological difficulties[8]. Here we present results of experiments in which track logs of homing pigeons flying in flocks of up to 10 individuals have been obtained by high-resolution lightweight GPS devices and analyzed using a variety of correlation functions inspired by approaches common in statistical physics. We find a well-defined hierarchy among flock members from data concerning leading roles in pairwise interactions, defined on the basis of characteristic delay times between birds' directional choices. The average spatial position of a pigeon within the flock strongly correlates with its place in the hierarchy, and birds respond more quickly to conspecifics perceived primarily through the left eye – both results revealing differential roles for birds that assume different positions with respect to flock-mates. From an evolutionary perspective, our results suggest that hierarchical organisation of group flight may be more efficient than an egalitarian one, at least for those flock sizes that permit regular pairwise interactions among group members, during which leader-follower relationships are consistently manifested.**

Collective movement phenomena in animals include many spectacular and familiar examples: among birds, seemingly instantaneous changes in a flock's direction of motion, the abrupt splitting of a flock, or a synchronised landing are all signs of rapid collective decision-making by group members, typically on a very short time scale. What behavioural rules govern such phenomena? The most elaborate way to address this question would be to obtain detailed spatiotemporal data on the positions of individuals during group movement. Nevertheless, up to now progress has been hampered by technological difficulties involved in tracking individuals with sufficiently high precision to resolve intra-group spatial relations in fast-moving animal collectives. As an alternative approach, numerous simulation models have been proposed to obtain insight into the basic laws of collective motion[3,9-11], yet rarely have detailed comparisons been attempted between these models and experimental data[7]. Outstanding questions include, whether, for example, all group members are "equal", as most models assume for the sake of simplicity, or whether one or a small number of leaders are able to contribute with differential influence to the group's movement decisions[12,13].

Over the last decade, rapid progress in sensor technology has enabled increasingly accurate tracking of free-flying birds, leading to important advances in our understanding of avian orientation strategies[14-17]. Applying advanced technologies to multiple individuals travelling as a group now also provides a novel window onto the rules underlying collective motion[18-22]. In particular, a new generation of GPS devices – capable of capturing movement decisions at the scale of a fraction of a second – allow us to make use of sophisticated evaluation techniques for exploring the influence that individual group members have on a fast-moving collective's behaviour. We used a combination of state-of-the-art GPS loggers with quantitative methods inspired by statistical physics to produce a detailed mapping of individual directional choice dynamics and potential leading activity within flocks of up to 10 homing pigeons (Supplementary Fig. 1).

We recorded the birds' movement under two conditions: while the flock was engaged in spontaneous flights near the home loft ("free flights") and during homing following ~15-km displacement from the loft ("homing flights"; see Supplementary Fig. 2). To investigate the influence that a given bird's behaviour had on its fellow flock members and on the flock as a whole, we evaluated the temporal relationship between the bird's flight direction and those of others (Fig. 1). A leading event was said to have occurred when a bird's direction of motion was "copied" by another bird delayed in time. To quantify such effects we determined the directional correlation delay time $\tau_{ij}^*$ for each pair of birds $i$ and $j$ (see Fig. 1c and Supplementary Methods for further detail). Then, from the pairwise $\tau_{ij}^*$ values, we composed a directional leader-follower network for each flight. In such a network the nodes represent individual birds, while the edges (links) denote inferred relations between their movements. We constructed networks by including only those edges whose directional correlation values based on $\tau_{ij}^*$ were above a given variable minimum, $C_{\min}$. The resultant networks were then quantified in terms of the degree of hierarchical organization they exhibited.

We concentrated on analysing velocity correlations because of the well-supported assumption that information obtainable from spatiotemporal functions has considerably better accuracy than steady global positional data. Since we calculate, e.g., the directional correlation delay data from long series of smoothly changing trajectories averaged over a large number of point pairs, most of the noise will average out. In addition, we found that our GPS devices reproduced *shifts* in the direction of motion much more accurately than global position itself. Thus, quantities based on the interrelations of the derivatives of the trajectories suffer from significantly less uncertainty. We have verified the validity of this assumption quantitatively by generating sample trajectories with given superimposed positional perturbations (see Supplementary Methods).

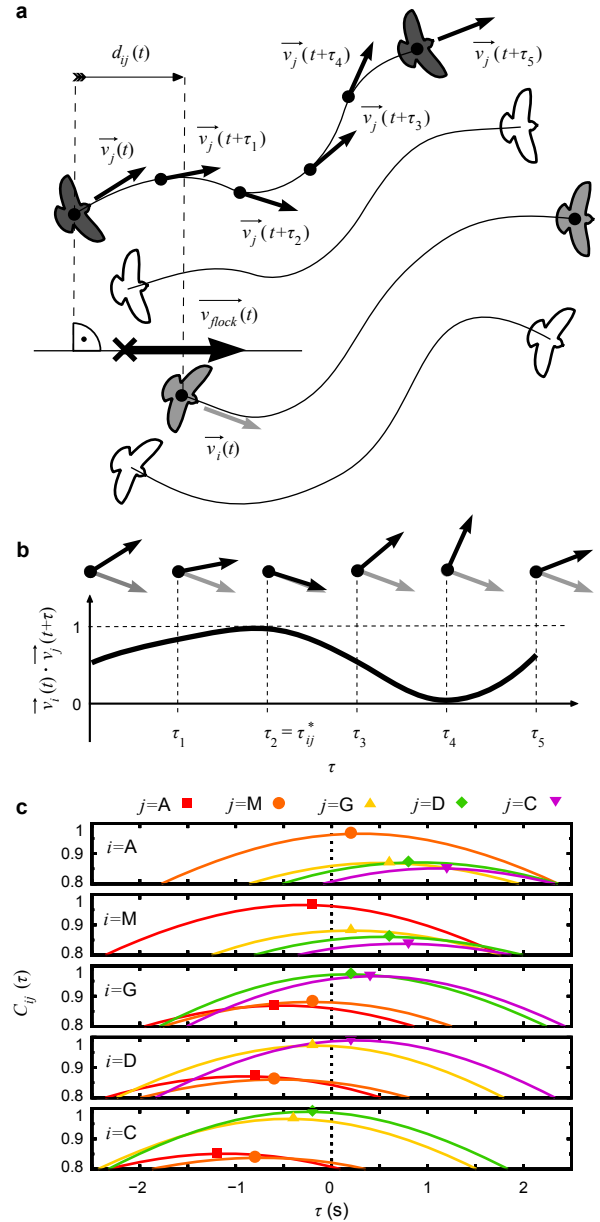

**Figure 1. Summary of directional correlation function analysis for determining leader-follower relationships within a flock. a,** Method for determining $d_{ij}(t)$, the projected distance of birds $i$ (light grey) and $j$ (dark grey) onto the direction of motion of the whole flock at each time step, $t$. The cross indicates the center of mass of the flock. $\vec{x}_i(t) - \vec{x}_j(t)$, the relative position of the birds, is projected onto $\vec{v}_{flock}(t)$, the average velocity of the whole flock. For each pair ($i \neq j$) the directional correlation function is $C_{ij}(\tau) = \langle \vec{v}_i(t) \cdot \vec{v}_j(t+\tau) \rangle$, where $\langle ... \rangle$ denotes time average. The arrows show the direction of motion, $\vec{v}_i(t)$. **b,** Visualization of scalar product of the normalized velocity of bird $i$ at time $t$ and that of bird $j$ at time $t+\tau$ in panel (a). Here, bird $j$ is following bird $i$ with correlation time $\tau_{ij}^*$. **c,** The directional correlation function $C_{ij}(\tau)$ during a flock flight (that shown in Fig. 2). For more transparency only the data of birds A, M, G, D and C (in the order of hierarchy for that flight) are shown. The solid symbols indicate the maximum value of the correlation function, $\tau_{ij}^*$.

About two-thirds (63%) of pairwise comparisons between birds of a flock produced clearly directed edges ($C_{min}$=0.5). That is, birds tended to copy consistently the directional behaviour of particular individuals, while being copied in their orientational choices by others. The average directional correlation delay time was 0.37 s (± 0.27 s SD) for $C_{min}$=0.5 and 0.32 s (± 0.20 s SD) for $C_{min}$=0.9. Such characteristic delay times can thus be taken to represent birds' reaction times in the context of following a persistent change in the direction of motion of neighbouring birds (rather than, for instance, the considerably shorter reflex-like reactions of a startle response[23]).

Crucially, most flights produced a robust hierarchical network (see Fig. 2 for an example), containing only transitive leader-follower relationships. Only 3 of 15 flights contained directed loops within the network, and across all flights, the proportion of the total number of edges which pointed in the same direction averaged 0.99 (± 0.03 SD) (Supplementary Table 1). Furthermore, randomization tests suggest that the probabilities of obtaining by chance networks with as many or fewer loops as those we observed are extremely low (Erdős-Rényi model for random directed networks, p < 0.001; Supplementary Table 1). Hierarchically organised group movement thus appears to be a reliably observable, robust phenomenon in pigeon flocks of the sizes we tested (up to 10 individuals) – opening up a suite of important questions about the roles, identities, and benefits accrued by members that assume the relative ranks of leaders and followers.

Do, for example, leader-follower relationships within specific pairwise comparisons extend across multiple flights? We calculated the average directional correlation delay times, $\tau_{ij}^*$, for all pairs who flew together on at least two occasions and for whom $C_{min}=0.99$. The overall network thus composed was also hierarchical, containing 9 nodes and 24 edges (Fig. 3a). In addition, we examined the effect of individual birds on the movement of the group as a whole, by assessing the average directional correlation delay time for every bird and the rest of the flock. This measure, denoted $\overline{\tau}_i$, allows us in turn to fully resolve hierarchical order among all nine birds, by creating a linear ranking consistent with all available data on edges (see also Supplementary Figures 3 and 4). The perfect correspondence between the order of $\overline{\tau}_i$ values and hierarchical rank (allowing for relative rankings that cannot be decided on the basis of edges alone; Fig. 3a) confirms that birds higher in the hierarchy were more influential in determining the direction of the flock's movement. This finding provides powerful support for our conclusion that certain individuals are able to contribute with relatively more weight to the movement decisions of the flock, through having followers who consistently copy their movement. We note that $\overline{\tau}_i$ values obtained separately for free and homing flights correlate significantly (Pearson's r = 0.797, n = 8, p = 0.018), suggesting that certain birds have a propensity to act as leaders irrespective of navigational context.

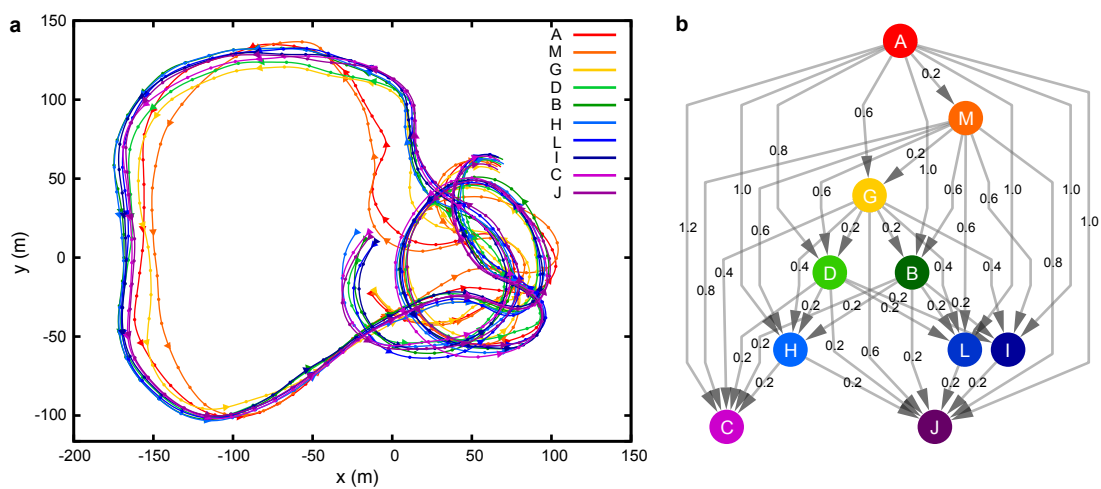

**Figure 2. Hierarchical leadership network generated for a single flock flight**. **a,** 2-minute segment from a free flight performed by a flock of ten pigeons. Dots and triangles indicate every 1s and 5s, respectively; triangles point in the direction of motion. Letters refer to bird identity. **b,** Hierarchical network of the flock for the flight shown in (a). For each pairwise comparison the directed edge points from the leader to the follower; values on edges show the time delay (in seconds) in the two birds' motion. For pairs of birds not connected by edges directionality could not be resolved at $C_{min}$= 0.5.

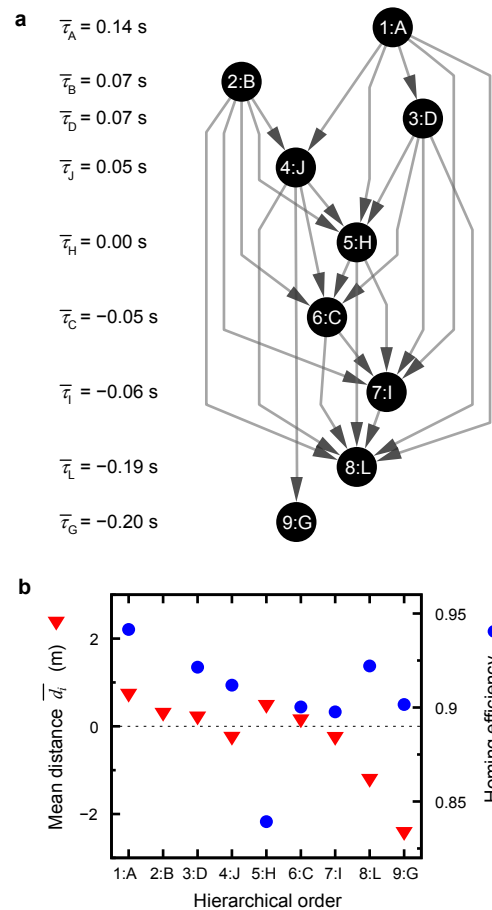

Intuitively, we expect individuals near the front of the group to be responsible for the majority of directional decisions, and evidence from a variety of species confirms that this is a reasonable assumption[24,25]. Nevertheless, in flying birds, with a field of vision close to 340° which allows individuals to track the movements of those behind them, the assumption is less trivial. We therefore determined for each bird its average distance from the centre of the flock projected onto the direction of motion of the flock, $\overline{d}_i$. We found a strong correlation between $\overline{d}_i$ and overall hierarchical order (red symbols in Fig. 3b; Pearson's correlation for $\overline{d}_i$ vs. $\overline{\tau}_i$, r = 0.863, n = 9, p = 0.003), which supports the notion that individuals occupying positions near the front of the flock tend also to assume leadership roles (see also Supplementary Movies 1 and 2).

Interestingly, besides the front-back distinction between leaders and followers, we also found evidence of a left-right effect. During homing, the more time a bird spent behind a particular partner, the more likely it was to be flying to that partner's right (and would thus have been perceiving it predominantly through its left eye; Table 1). Birds' visual systems are known to be lateralised[26], with a superiority of the left brain hemisphere (which receives input contralaterally, from the right eye) in large-scale spatial tasks[27], and a right-hemispheric (left-eye) specialisation for social input (such as individual recognition[28]). Accordingly, our data also indicate that when birds perceive a particular partner predominantly through the left eye they respond more quickly and/or strongly to its movements (Table 1) suggesting that social information may be preferentially processed through the left-eye/right-hemispheric system.

**Figure 3. Hierarchical leadership network generated from multiple flock flights. a,** Overall hierarchical network of all birds that flew together on at least two occasions ($C_{min}$= 0.99). The flock-averaged directional correlation delay time for each bird, $\overline{\tau}_i$, is indicated on the left; note that it has the same order as the network, as it was used to order those birds between whom relative ranks could not be resolved on the basis of edges alone. **b,** Average projected distance onto the direction of motion of the flock, $\overline{d}_i$ (red triangles), and solo homing efficiency (beeline distance / distance travelled; blue circles) as a function of the hierarchical order resolved in (a). Due to GPS logger failure, solo efficiency data is missing for Bird B.

**Table 1. Analysis of laterality effects during group homing flights.**

| Flight number | n | $Q_{left}$ vs. $Q_{forward}$ Pearson corr. r | p | $Q_{left}$ vs. $d_{ij}$ Pearson corr. r | p | $\tau_{left} - \tau_{right}$ Mean (s) | s.d. (s) | Student-t test* t-value | p |
|---|---|---|---|---|---|---|---|---|---|
| 1 | 90 | 0.37 | **<0.001** | 0.32 | **0.002** | -0.23 | 0.27 | -8.03 | **<0.001** |
| 2 | 72 | 0.23 | **0.048** | 0.25 | **0.036** | -0.19 | 0.21 | -7.72 | **<0.001** |
| 3 | 46 | 0.59 | **<0.001** | 0.62 | **<0.001** | -0.016 | 0.026 | -4.39 | **<0.001** |
| 4 | 72 | 0.49 | **<0.001** | 0.54 | **<0.001** | -0.006 | 0.020 | -2.32 | **0.023** |

n: Number of data pairs for given flight. The total number represents all possible pairwise comparisons between birds of the flock. For each pair, only those datapoints were analysed where the two individuals were < 10 m apart (see Supplementary Methods). Note that during flight #3, two birds broke away from the group soon after release, and did not have sufficient data at the given distance limit for comparisons with every other flock-mate.

$Q_{left} = t_{left} / t_{total}$ : Left Ratio. For any given pair, time spent with partner positioned on focal bird's left (relative to its direction of motion) divided by the total time spent flying together.

$Q_{forward} = t_{forward} / t_{total}$ : Forward Ratio. For any given pair, time spent by partner ahead of focal bird (relative to the direction of motion of the whole flock) divided by the total time spent flying together (see also Supplementary Fig. 6).

$d_{ij}$ : Average projected distance onto the direction of motion of the whole flock for each pair.

$\tau_{left} - \tau_{right}$ : Difference of the $\tau_{left}$ and $\tau_{right}$ values for any given pair, where $\tau_{left}$ and $\tau_{right}$ refer to directional correlation delay times calculated separately for datapoints where the partner is positioned to the left and to the right of the focal bird, respectively.

* The Student t value is calculated on the basis of the distribution of $\tau_{left} - \tau_{right}$ values obtained when the observed $\tau_{left}$ and $\tau_{right}$ pairs are randomly reassigned into novel pairings, and thus tests whether within-bird observed differences in directional correlation delay times are significantly different from the random expectation. In all four flights the mean is significantly lower than 0, suggesting that birds respond faster to their partners when the latter are in view primarily of the left eye.

To explore whether a bird's propensity to lead relates to individual navigational performance, we conducted a single solo homing test, releasing individually the nine subjects represented in the overall hierarchy of Figure 3a. One bird ("H") flew a considerably longer path than the average for the remaining subjects (> mean + 5 SD); when this outlier is excluded, the correlation between leadership rank and homing efficiency approaches significance (Pearson's r = -0.71, n = 7, p = 0.074; blue symbols in Fig. 3b) although not if it is included (Pearson's r = -0.29, n = 8, p > 0.100). Thus, although the current data are equivocal, they are suggestive that leadership may be related to individual navigational efficiency, with birds higher in the hierarchy also demonstrating more accurate solo navigation. Whether such effects would derive from more motivated or inherently better navigators being better able to assume leadership roles[13], or from birds that have had more experience leading also having had increased opportunities for navigational learning (the passenger/driver effect[29]) remains an intriguing open question regarding the causes – or indeed consequences – of leadership.

Could the mechanism we identified in small pigeon flocks scale up to larger groups? If hierarchies can operate at multiple levels (as they do during group movement by, e.g., zebra herds[25]), then it is conceivable that much larger collectives can rely on recursion in the decision-making process. In human collectives complex hierarchical structures are widespread – although these tend to reflect deliberate organisation and contain fixed roles for individuals. In the case of our flocks, pairwise leader-follower relations may be established more spontaneously, in a state-dependent fashion[30], based on individuals' current motivation, navigational knowledge, ability, and so forth. Since these attributes can vary over short time-scales, individuals' roles are manifested in a dynamically changing manner, i.e., only in average, with the leading role of a given bird fluctuating over time. Our quantitative results reveal a delicate arrangement of these dynamic leader-follower relations into a hierarchical network comprising a spectrum in levels of leadership; a sophisticated system that may, from an evolutionary perspective, bring benefits to individual group members over, for example, a single-leader scenario, or an ancestrally (presumably) egalitarian collective. Finally, because of the general nature of the consensus-finding mechanism we observed, it has the potential of being applicable to a wider range of collectives as well.

## Methods

**Subjects and experimental protocols.** 13 homing pigeons, all aged between 1 and 5 years, participated in the experiments. All had had previous homing experience and most had previously competed in races (>100 km) for young pigeons. Birds were habitually allowed to fly freely outside the loft twice a day. All subjects (labelled A to M) were initially equipped with plasticine dummy weights (16 g, same size and weight as the GPS logger), affixed to the back with an elastic harness, to habituate them to flying and living with a load. We collected GPS data from three types of releases: free flights of flocks around the home loft (11 flights in total; with flocks spending on average 12 min in the air), homing flights in flocks (4 flights; all participating subjects released simultaneously), and individual homing flights (one per subject). Group homing flights were conducted from release sites located 13.7-14.8 km from the loft; the single solo flight from 15.2 km (600 m from one of the sites used during group releases). The different types of flights were interspersed in the following order: 1 free, 1 flock homing, 1 free, 3 flock homing, 1 individual homing, and 9 free. In most cases, flocks consisted of 10 (8 flights) or 9 (5 flights) pigeons, while on two occasions the flock numbered 8 individuals, and once only 7 participated. A maximum of two flights were conducted per day, between 22nd of August and 26th of September 2008. In total, GPS devices logged 32 h of flight time, representing 580,000 datapoints gathered for analysis.

**GPS device and data handling.** The GPS device we developed was based on a commercially available U-blox (Thalwil, Switzerland) product. It was capable of logging 30,000 datapoints (latitude, longitude, and altitude coordinates and time), measured 2.5 x 4.5 cm, and weighed 16 g (3-4% of the subjects' body weight). The temporal resolution of the device was 0.2 s. Immediately before recorded flights the dummy was replaced by the GPS device, and upon recapture of the birds at the loft the device was removed and the log files downloaded to a computer. The geodetic coordinates provided by the GPS were converted into x, y, and z coordinates using the Flat Earth model. These coordinates were smoothed by a Gaussian filter (σ = 0.4 s), and the cubic B-Spline method was used to fit curves onto the points obtained with the 0.2 s sampling rate. Occasionally, the device failed to log every second or third point; in such cases we interpolated the position of the missing datapoints by averaging those recorded immediately before and after. As with the GPS measurement the error of the z coordinate is much larger than that in the horizontal directions, and because the birds' main movement decisions could be projected onto the horizontal plane, we used only x and y in our analysis. In independent tests we confirmed that the accuracy of the x and y global coordinates was in the range of 1-2 m. While this degree of accuracy does not permit accurate determination of spatiotemporal configurations of individuals within the flock, it is nevertheless sufficient for calculating various relevant correlation functions that characterise relations among the birds' motion (see Supplementary Methods).

**Supplementary Information** and **2 Supplementary Movies** accompany the paper on **http://hal.elte.hu/pigeonflocks**.

**Acknowledgements**. We thank the Hungarian Racing Pigeon Sports Federation – in particular its president, I. Bárdos – for their support. We are grateful to J. Pató for allowing his pigeons to participate in the present research, and for his technical help throughout the experiments. This research was partially supported by the EU ERC COLLMOT and the EU FP6 STARFLAG projects. DB is grateful to the Royal Society and to Somerville College, Oxford for financial support.

**Author Contributions.** Zs. A. and T.V. designed the experiments. Zs. A. and M.N. performed the experiments, M.N and D.B. designed the evaluation of data. M.N. performed the analysis and the visualization of data. D.B. and T.V wrote the paper.

**Author Information.** Correspondence and requests for materials should be addressed to T. V. (vicsek@hal.elte.hu).